\documentclass[aps,prl,showpacs,showkeys,twocolumn,groupedaddress]{revtex4}
\usepackage{graphics}
\usepackage{epsfig}

\begin{document}

\title{Magnetovolume instabilities in the pressure dependence of the K-edge circular dichroism of Fe$_{3}$C Invar particles}

\author{E. Duman}
\author{M. Acet}
\altaffiliation{Corresponding Author, Electronic address :
macet@ttphysik.uni-duisburg.de, tel.: ++49 (203)379 2023 fax: ++49
(203)379 2098}
\author{E. F. Wassermann}
\affiliation{Experimentalphysik, Universit\"{a}t Duisburg-Essen,
D-47048 Duisburg, Germany}

\author{J. P. Iti\'{e}}
\affiliation{Physique des Milieux Condensés, CNRS-UMR 7602,
Université Paris VI, B77 4, place Jussieu, F-75252 Paris CEDEX 05,
France}

\author{F. Baudelet}
\affiliation{Syncrotron Soleil L'Orme des Merisiers,
Saint-Aubin-BP 48 F-91192 Gif-sur-Yvette Cedex France}

\author{O. Mathon and S. Pascarelli}
\affiliation{European Synchrotron Radiation Facility, B.P. 220,
F-38043 Grenoble Cedex, France}

\date{\today}

\begin{abstract}
The p-electrons of carbon in the interstitial compound Fe$_{3}$C
hybridize with the Fe d-band and enhance the valence electron
concentration of Fe from 8 to 8.67. At this concentration,
substitutional 3d transition metals and alloys exhibit strong
moment-volume coupling phenomena and associated magnetovolume
instabilities, otherwise known as the Invar effect. For this
reason, Fe$_{3}$C is also expected to incorporate a strong
magnetovolume instability, and therefore, we examine the pressure
dependence of the of the K edge x-ray magnetic circular dichroism
in Fe$_{3}$C at ambient temperature and pressures up to 20 GPa. We
find clear evidence for a high-moment to low-moment transition at
about 10 GPa.
\end{abstract}



\maketitle

Metalloid atoms such as boron, carbon or nitrogen take up
interstitial lattice positions in 3d metals when forming alloys
and compounds. The outer p-states of the metalloid atoms hybridize
with the host metal d-states in such a manner that the effective
electron concentration of the metal increases \cite{Acet}. Fe has
a valence electron concentration ${e/a}$ of 8 (s+d) electrons per
atom, whereas in Fe$_{3}$C, in which C has two p electrons, the
electron concentration per metal atom becomes
{[8(0.75)+2(0.25)]/0.75}=8.67. This concentration corresponds to a
value where large spontaneous magnetostriction associated with
magnetovolume instabilities in the alloys of 3d systems are found.
Such magnetostrictive phenomena are known as the Invar effect, and
the archetype material is the face centered cubic
Fe$_{0.65}$Ni$_{0.35}$ Invar alloy with (${e/a}$)=8.66
\cite{Wassermann}.

A magnetovolume instability is a rapid change of the magnetic
moment with respect to a small change in the atomic volume. Under
applied pressure at low temperature, a system incorporating such
an instability can undergo a transition from a large-volume
high-moment state (about 1.7 $\mu_{B}$ for Fe$_{0.65}$Ni$_{0.35}$)
to a small-volume low-moment state of nearly vanishing magnetic
moment. These states are separated by a small energy difference of
several meV \cite{Moruzzi}, which is a value within the thermal
range of the solid state. Therefore, it is thought that by
increasing the temperature, the small-volume low-moment state is
progressively occupied at the expense of the large-volume
high-moment state causing a contraction that counteracts the
normal lattice expansion. The overall effect is then a nearly
vanishing thermal expansion coefficient in a broad temperature
range; namely, the Invar effect.

Differing from the structure of Fe$_{0.65}$Ni$_{0.35}$, Fe$_{3}$C
is orthorhombic, and the Fe atoms are arranged in a manner such
that they have two different environments denoted as FeI and FeII
\cite{Pepperhof}. However, both materials have nearly equivalent
electron concentrations and exhibit nearly identical Invar-typical
features in the temperature dependence of the thermal expansion
\cite{Pepperhof} and the bulk modulus \cite{Duman}, although
Fe$_{3}$C consists of only a single metal atom species. Such data
are supporting evidence for the presence of the Invar effect in
Fe$_{3}$C, however, they provide information only on the lattice
properties, and do not give evidence for the presence of
instabilities in the magnetic degrees of freedom. This evidence is
necessary to understand that the spontaneous volume enhancements
and the associated lattice anomalies in Fe$_{3}$C are related to
its magnetism.

The direct method of detecting a presence of a magnetic
instability that is coupled to the lattice degrees of freedom is
to measure a magnetization related parameter under applied
pressure. This may be done using techniques such as M\"{o}ssbauer
spectroscopy \cite{Elmeguid89}, ac-susceptibility
\cite{Matsushita}, x-ray magnetic circular dichroism (XMCD)
\cite{Odin}, or x-ray emission spectroscopy \cite{Rueff}, usually
with the sample located in a diamond anvil cell (DAC). For the
case of Fe$_{3}$C, the M\"{o}ssbauer technique has the drawback
that it requires an $^{57}$Fe enriched sample because of the small
quantity of sample material that can be brought between the
diamond anvils. Unrealistic quantities of $^{57}$Fe would be
required to prepare a sample by chemical separation from a
starting Fe+Fe$_{3}$C ingot, or by any dissociation process of
Fe(CO)$_{5}$. The ac-susceptibility technique requires macroscopic
size samples in the order of a cubic millimeter, and Fe$_{3}$C
cannot be sintered in pure form, since it decomposes under
pressure at the required high sintering temperatures. The XMCD
technique with Fe$_{3}$C under pressure in a DAC, on the other
hand, does not have any of the drawbacks mentioned above, and can
be applied to observe variations in the magnetic degrees of
freedom with applied pressure. Nevertheless, due to the strong
absorption of the diamonds at low x-ray energies corresponding to
the Fe L-edge, for which quantitative analysis has a chance, one
must work at the alternative Fe K-edge, for which the results can
only be interpreted qualitatively at present due to complexities
involved in theoretical modeling.

\begin{figure}[t]
\includegraphics[width=8 cm]{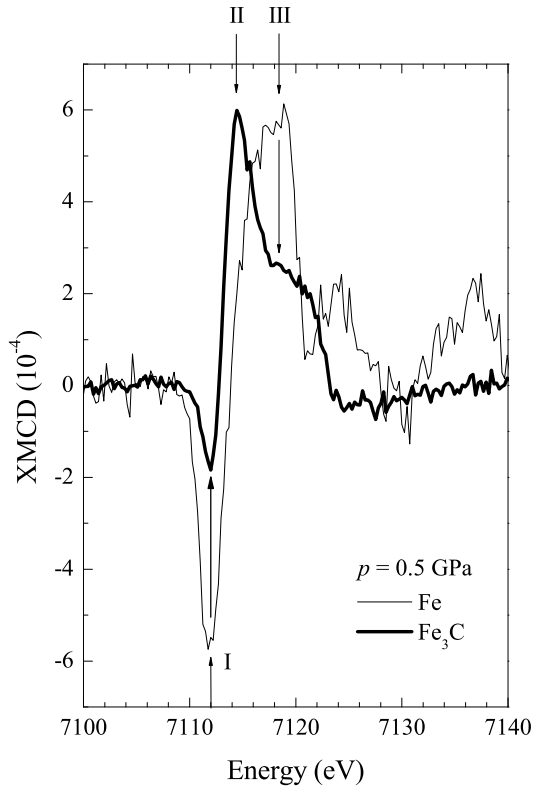}
\caption{\label{Figure1}XMCD of Fe$_{3}$C (heavy line) and Fe
around the Fe K-edge.}
\end{figure}

We have investigated the pressure dependence of the XMCD at the
K-edge of Fe in Fe$_{3}$C at room temperature and pressures up to
20 GPa on increasing and decreasing pressure. The measurements
were carried out at the ESRF on the ID24 beam line installed on an
undulator source. Circular polarization was attained using a
quarter wave plate. Two sets of measurements at each pressure with
both polarization ellipticities and both magnetic field directions
were taken in order to eliminate systematic errors arising from
the flipping of the quarter wave plate and of the magnetic field.
A magnetic field of 0.4 T was applied with an electromagnet. The
magnetization in this field reaches about 80 \% of the saturation
value of the sample used in the present experiments \cite{Hulser}.
Fe$_{3}$C particles of about 50 nm were prepared by the
dissociation of Fe(CO)$_5$ in a hot wall reactor in the presence
of methane and were separated from the admixture of C and Fe by
chemical and physical methods \cite{Duman}. After separation,
their purity, morphology and structure were checked by electron
microscopy and x-ray analysis. No foreign phase or any mixture
with Fe was detected.

\begin{figure}[t]
\includegraphics[width=8 cm]{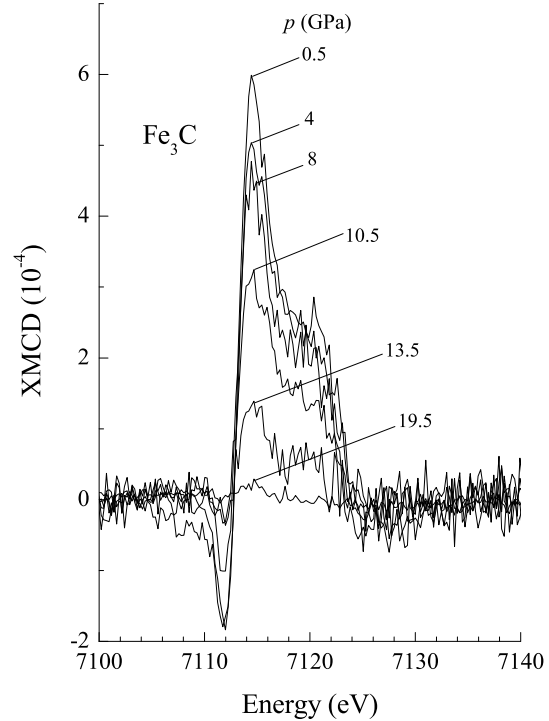}
\caption{\label{Figure2}XMCD of Fe$_{3}$C around the Fe K-edge at
selected pressures.}
\end{figure}

Figure 1 shows a comparison of the K-edge XMCD spectra of Fe and
Fe$_{3}$C with both data obtained in the diamond anvil cell under
a loading pressure of 0.5 GPa. The present data for Fe are similar
to data at ambient pressure previously reported \cite{Mathon}. The
position of the first dip on the low energy side of both Fe and
Fe$_{3}$C lies at about 7112 eV (position I). The feature at
position II in the Fe$_{3}$C spectrum has no counterpart feature
in the Fe spectrum. The broad maximum in the Fe spectrum around
7118 eV coincides with the position around the shoulder in the
Fe$_{3}$C data at the position denoted as III. Since in Fe$_{3}$C,
Fe occupies two electronically nonequivalent sites, the occurrence
of features different from those in the Fe spectrum can be
attributed to the different interactions of the excited 4p
photoelectrons with the spin polarized d-bands for the two
different Fe sites in Fe$_{3}$C.

Figure 2 shows the XMCD spectra from ambient pressure up to about
20 GPa taken on increasing pressure. The data on decreasing
pressure are similar and are not shown here. The overall intensity
of the the spectra diminishes as the pressure increases, and only
a remanent feature remains at 19.5 GPa.

\begin{figure}[t]
\includegraphics[width=8 cm]{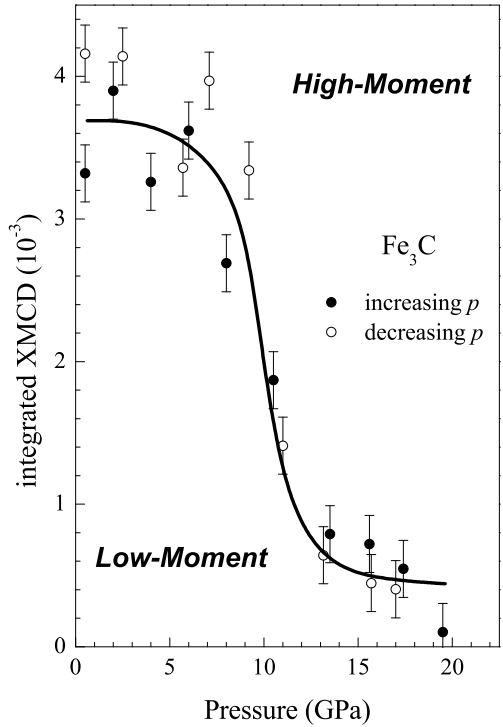}
\caption{\label{Figure3}Integrated XMCD of Fe$_{3}$C. The
high-moment to low-moment transition takes place at about 10 GPa.}
\end{figure}

The integrated XMCD intensity, obtained after subtracting the
background intensity before and after the K-edge, is plotted in
Fig. 3. A clear transition from a high-moment to a low-moment
state is observed without any hysteresis in the increasing and
decreasing pressure data. The intensity initially shows no
appreciable variation with pressure and then begins to decrease
rapidly around 8 GPa. At about 13 GPa, it has dropped down to
about 80 \% of its value at ambient pressure. Assuming a direct
proportionality between the integrated XMCD and the average
magnetic moment, the moment of 1.8 $\mu_{B}$ in the ground state
of Fe$_{3}$C (high-moment state) \cite{Hofer, Shabanova73,
Shabanova75, Haglund} can be estimated to drop down to about 0.4
$\mu_{B}$. At about 12 GPa, the lattice constants acquire the
values ${a}$=0.496 nm, ${b}$=0.662 nm, and ${c}$=0.440 nm and the
cell volume is ${V}$=0.144 nm$^{3}$ \cite{Duman}. Under ambient
conditions, ${a}$=0.504 nm, ${b}$=0.673 nm, and ${c}$=0.448 and
the equilibrium cell volume is ${V}_0$=0.151 nm$^{3}$. Therefore,
a change in atomic volume of about 5 \% is required to induce the
transition. This amount is about the same as for
Fe$_{0.65}$Ni$_{0.35}$ \cite{Elmeguid89}. However, the pressure
required to induce the transition is only about 5 GPa for
Fe$_{0.65}$Ni$_{0.35}$ because, it is softer with a bulk modulus
of 130 GPa \cite{Renaud} as opposed to 174 GPa \cite{Duman}for
cementite.

By probing the magnetic degree of freedom with Fe K-edge XMCD
spectroscopy, we find direct evidence for the presence of a
magnetovolume instability in Fe$_{3}$C. These instabilities are
expected to be the source of the Invar-typical features observed
in the temperature dependence of the thermodynamical parameters
for this material. To gain more information on the effect of the
magnetism of the individual Fe sites in Fe$_{3}$C on the
magnetovolume instability, it is necessary to provide more
elaborate modeling for K-edge XMCD spectroscopy of Fe and Fe$_3$C.

\begin{acknowledgments}
This work was supported by the ESRF and Deutsche Forschungsgemeinschaft (SFB 445).
\end{acknowledgments}



\begin{thebibliography}{99}
\bibitem{Acet} M. Acet, B. Gehrmann, E. F.Wassermann, H. Bach and W. Pepperhoff, J. Magn. Magn. Mater. \textbf{232}, 221 (2001).

\bibitem{Wassermann} E. F. Wassermann, in Ferromagnetic Materials, Ed. K. H. J. Buschow and E. P. Wohlfarth (North Holland, Amsterdam, 1990) Vol. 5, S. 238.

\bibitem{Moruzzi} V. L. Moruzzi, Phys. Rev. B, \textbf{40}, 6939, (1990).

\bibitem{Pepperhof} W. Pepperhof and M. Acet, Constitution and Magnetism of
Iron and its Alloys, (Springer-Verlag, Berlin-Heidelberg, 2001),
p.166.

\bibitem{Duman} E. Duman, M. Acet, T. H\"{u}lser, E.F. Wassermann, B. Rellinghaus,
J.P. Iti\'{e} and P. Munsch, J. Appl. Phys., (to be published).

\bibitem{Elmeguid89} M. M. Abd-Elmeguid and H. Micklitz, Physica B, \textbf{161}, 17 (1989).

\bibitem{Matsushita} M. Matsushita, S. Endo, K. Miura, and F. Ono, J. Magn. Magn.
Mater. \textbf{265}, 352 (2003).

\bibitem{Odin} S. Odin, F. Baudelet, J. P. Itié, A. Polian, S. Pizzini, A.
Fontaine, Ch. Giorgetti, E. Dartyge and J. P. Kappler, J. Appl.
Phys., \textbf{83}, 7291 (1998).

\bibitem{Rueff} J. P. Rueff, A. Shukla, A. Kaprolat, M. Krisch, M. Lorenzen, F.
Sette and R. Verbeni, Phys. Rev. B, \textbf{63}, 132409 (2001).

\bibitem{Hulser} T. H\"{u}lser, unpublished

\bibitem{Mathon} O. Mathon F. Baudelet, J. P. Iti\`{e}, S. Pasternak, A. Polian and S. Pascarelli, J. Synchr.
Rad., \textbf{11}, 423, (2004)

\bibitem{Hofer} L. J. E. Hofer and E. M. Cohn, J. Am. Chem. Soc. \textbf{81}, 1576 (1959).

\bibitem{Shabanova73} I. N. Shabanova and V. A. Trapeznikov, JETP Letters, \textbf{18}, 339,
(1973).

\bibitem{Shabanova75} I. N. Shabanova and V. A. Trapeznikov, J. Electron Spectrosc.
Relat. Phenom, \textbf{6}, 297, (1975).

\bibitem{Haglund} J. H\"{a}glund, G. Grimvall and T. Jarlborg, Phys. Rev. B,
\textbf{44}, 2914 (1991).

\bibitem{Renaud} P. Renaud and S. G. Steinemann, Physica B, \textbf{161}, 75, (1989).

\end{thebibliography}
\end{document}